\documentclass[aps,prd,twocolumn,showpacs,superscriptaddress,groupedaddress]{revtex4}
\usepackage{amsmath}
\usepackage{latexsym}
\usepackage{amsfonts}
\usepackage{graphicx}
\usepackage{subfig}
\usepackage{mathrsfs}
\usepackage{epstopdf}
\usepackage{ulem}
\usepackage{bm}
\usepackage{color}
\usepackage[colorlinks,breaklinks=true,linkcolor=blue,citecolor=blue]{hyperref}
\usepackage{multirow}
\usepackage{footnote}
\usepackage{url}

\newcommand{\br}{\mathcal{R}}

\begin{document}

\title{Relieving the Tension between Weak Lensing and Cosmic Microwave Background with Interacting Dark Matter and Dark Energy Models}

\author{Rui An}
\email{an\_rui@sjtu.edu.cn}
\affiliation{IFSA Collaborative Innovation Center, School of Physics and Astronomy, Shanghai Jiao Tong University, Shanghai 200240, China}

\author{Chang Feng}
\email{changf@illinois.edu}
\affiliation{Department of Physics, University of Illinois at Urbana-Champaign, 1110 W Green St, Urbana, IL, 61801, USA}

\author{Bin Wang}
\email{wang\_b@sjtu.edu.cn}
\thanks{Corresponding Author}
\affiliation{Center for Gravitation and Cosmology, College of Physical Science and Technology, Yangzhou University, Yangzhou 225009, China}
\affiliation{IFSA Collaborative Innovation Center, School of Physics and Astronomy, Shanghai Jiao Tong University, Shanghai 200240, China}

\begin{abstract}
We constrain interacting dark matter and dark energy (IDMDE) models using a 450-degree-square cosmic shear data from the Kilo Degree Survey (KiDS) and the angular power spectra from Planck's latest cosmic microwave background measurements. We revisit the discordance problem in the standard Lambda cold dark matter ($\Lambda$CDM) model between weak lensing and Planck datasets and extend the discussion by introducing interacting dark sectors. The IDMDE models are found to be able to alleviate the discordance between KiDS and Planck as previously inferred from the $\Lambda$CDM model, and moderately favored by a combination of the two datasets.

\end{abstract}

\maketitle

\section{Introduction}

Cosmic shear, measured from distorted images of distant galaxies, can effectively map a three-dimensional dark matter structure in the late universe, making it a sensitive probe to constraining cosmological models. Recently, the standard Lambda cold dark matter ($\Lambda$CDM) model was examined  by employing weak lensing data taken from a $450$-$\text{deg}^2$ observing field of the Kilo Degree Survey (KiDS)~\cite{Hildebrandt2017} and it was disclosed that there exists a ``substantial discordance'' inferred from the $\Lambda$CDM model between the KiDS data \cite{Jong2015,Kuijken2015,Hildebrandt2017,Fenech2017} and the Planck 2015 cosmic microwave background (CMB) data \cite{Ade2016,Aghanim2016}. The discordance is at the level of $2.3\sigma$ and it was argued that this discordance can not be resolved even after reducing systematic uncertainties \cite{Hildebrandt2017,Joudaki2017}. In addition to the $\Lambda$CDM discordance between the KiDS and Planck datasets, earlier  similar tension was also found to be substantial between the Canada-France-Hawaii Telescope Lensing Survey (CFHTLenS) \cite{Heymans2012,Hildebrandt2012,Erben2013,Miller2013} and Planck datasets~\cite{Ade2014,MacCrann2015,Ade2016,Kohlinger2016,Joudaki2016}.

Besides the weaking lensing--CMB discordance, the standard $\Lambda$CDM model is also challenged by other observations. There is approximately a $3\sigma$ tension in  cosmological parameter space when the $\Lambda$CDM model is compared to Planck CMB data and local measurements of the Hubble constant based on the cosmic distance ladder \cite{Riess2011,Riess2016}. Recently, the Baryon Oscillation Spectroscopic Survey (BOSS) experiment of the Sloan Digital Sky Survey (SDSS) showed new evidence against the standard $\Lambda$CDM model \cite{Delubac2015} using measured baryon acoustic oscillations (BAO) in flux correlation functions of the Lyman-$\alpha$ forest from $158, 401$
quasars at high redshifts $(2.1\leq z\leq 3.5)$. The results indicate a $2.5\sigma$ deviation from $\Lambda$CDM in the measurements of the Hubble constant and angular distance at an average redshift $z=2.34$.

The tensions between observations of large-scale structure and CMB measurements made by Planck motivated a lot of studies to extend the standard $\Lambda$CDM model. So far the weak lensing data from KiDS have been used to test the extended cosmological models, which include massive neutrinos, nonzero curvature, evolving dark energy, modified gravity and running of the scalar spectral index~\cite{Joudaki2017}. It was found that the discordance between KiDS and Planck can be alleviated by introducing an evolving dark energy equation of state. Also, the CFHTLenS-Planck discordance has been revisited by using the extended models with a sterile neutrino and scale-dependent equation of state~\cite{MacCrann2015,Kunz2015}. Again, it was found that the confidence contours between two datasets started to overlap, effectively relieving the discordance. For the discordance between local Hubble constant and Planck measurements, a model with nonstandard physics in  dark energy and dark radiation sectors was proposed to relieve the tension~\cite{Bernal2016, DiValentino2016, Riess2016}. Similarly, it was suggested that an extended model with dark sectors' interactions might be a solution to the discordance between the Hubble constant and the angular distance of BOSS at redshift $z=2.34$~\cite{Ferreira2014}.

Not only being challenged by observations, the standard $\Lambda$CDM model is also suffering theoretical problems, such as the cosmological constant problem \cite{Weinberg1989}, i.e., a disagreement between the observed value and the estimation of quantum field theory. Moreover, the cosmological constant of the $\Lambda$CDM  model can not explain why dark energy dominated evolution in the late universe, making a coincidence problem \cite{Chimento2003}. Given the fact that energy density fractions of dark matter and dark energy account for nearly $25\%$ and $70\%$ of the energy content of our universe, it is quite counter-intuitive to conceive that dark matter and dark energy co-exist independently in the course of cosmological evolution. From perspective of field theory, it is more natural to consider that there are interactions between dark matter and dark energy through either exotic energy or momentum transfer. And it was argued that a suitable amount of interactions between dark sectors can help alleviate the coincidence problem \cite{Amendola2000,Pavon2005,Boeher2008,Olivares2006,Chen2008,Ferreira2014}. Recently the models with interactions between dark sectors have been extensively tested with different observational datasets, such as measurements of the CMB and galaxy clusters, see the recent review \cite{Wang2016} and references therein for detailed descriptions of physical models and observations.

In this work, we will constrain the interacting dark matter and dark energy model (IDMDE) using the weak gravitational lensing data from KiDS and Planck CMB data from both temperature and polarization. Moreover, we will revisit the discordance problem between KiDS and Planck with the IDMDE model and check if it is favored as compared to the standard  $\Lambda$CDM model, and to what extent the tension would be relieved.

The paper is organized as follows. In Section II we introduce phenomenological models on the interactions between dark matter and dark energy with background dynamics equations and linear perturbations. In Section III we describe the KiDS measurements and statistical algorithms that are used to quantify tensions between different data sets.  In Section IV we compare predicted weak lensing tomographic band powers from the IDMDE models to the  KiDS-450 and Planck2015 datasets, and examine how the IDMDE models are favored by both KiDS and Planck data sets, and to what extent they alleviate the previous $\Lambda$CDM discordance problem. Finally, we discuss our results and conclude in Section V.

\section{Interacting dark matter and dark energy models}

In this section, we begin with the phenomenological interacting dark matter and dark energy models \cite{Wang2016} in the spatially flat Friedmann-Lemaitre-Robertson-Walker (FLRW) universe. In these models, the total energy density of dark sectors is conserved and energy densities of dark matter and dark energy evolve individually as
\begin{gather}
\label{eq.rhodm}
\dot{\rho}_{c}+3\mathcal{H}\rho_{c}=aQ,\\
\label{eq.rhode}
\dot{\rho}_{d}+3\mathcal{H}(1+w)\rho_{d}=-aQ,
\end{gather}
where $\mathcal{H}$ is the time-dependent Hubble constant defined by $\mathcal{H}=\dot{a}/a=aH$, $a$ is the scale factor, the dot $\dot{}$ is the derivative with respect to conformal time, and $w=P_d/\rho_d$ is the equation of state of dark energy. Here $Q$ denotes the interactions between dark sectors and it can be phenomenologically expressed as a linear combination of energy densities of dark matter and dark energy, i.e., $Q=3\lambda_{1}H\rho_{c}+3\lambda_{2}H\rho_{d}$ with $\lambda_1$ and $\lambda_2$ being free parameters describing interaction strength.  In Table~\ref{tab.model}, we list all of the phenomenological IDMDE models in which curvature perturbations are not divergent when conditions given by the last column are satisfied~\cite{He2009,Gavela2009}.

\begin{table}[ht]
\caption{Phenomenological interacting dark matter and dark energy models \label{tab.model}}
\begin{tabular}{cccc}
\hline
Model & $Q$ & $w$ & Constraints \\
\hline
I & $3\lambda_{2}H\rho_{d}$ & $-1<w<-1/3$ & $\lambda_{2}<0$ \\
II & $3\lambda_{2}H\rho_{d}$ & $w<-1$ & $0<\lambda_{2}<-2w\Omega_c$ \\
III & $3\lambda_{1}H\rho_{c}$ & $w<-1$ & $0<\lambda_{1}<-w/4$ \\
IV & $3\lambda H(\rho_{c}+\rho_{d})$ & $w<-1$ & $0<\lambda<-w/4$ \\
\hline
\end{tabular}
\end{table}

In the linear theory, equations of the first-order perturbations for dark matter and dark energy are given by \cite{Costa2013}
\begin{align}
\label{eq.deltac}
\dot{\delta}_c=&-(kv_c+\frac{\dot{h}}{2})+3\mathcal{H}\lambda_2\frac{1}{\br}(\delta_d-\delta_c),\\
\label{eq.deltad}
\dot{\delta}_d=&-(1+w)(kv_d+\frac{\dot{h}}{2})+3\mathcal{H}(w-c_e^2)\delta_d \notag\\
&+3\mathcal{H}\lambda_1\br(\delta_d-\delta_c) \notag\\
&-3\mathcal{H}(c_e^2-c_a^2)[3\mathcal{H}(1+w)+3\mathcal{H}(\lambda_1\br+\lambda_2)]\frac{v_d}{k},\\
\label{eq.vc}
\dot{v}_c=&-\mathcal{H}v_c-3\mathcal{H}(\lambda_1+\frac{1}{\br}\lambda_2)v_c,\\
\label{eq.vd}
\dot{v}_d=&-\mathcal{H}(1-3c_e^2)v_d \notag\\
&+\frac{3\mathcal{H}}{1+w}(1+c_e^2)(\lambda_1r+\lambda_2)v_d+\frac{kc_e^2}{1+w},
\end{align}
where $\delta_i=\delta\rho_i/\rho_i$ is the perturbed density contrast, $v_i$ is the peculiar velocity, and the subscript $i$ represents dark matter or dark energy. The variable $h=6\Theta$ is the synchronous gauge metric perturbation, $\Theta$ describes a small deviation from a homogeneous and isotropic universe, $c_e$ is the effective sound speed of dark energy which is set to $1$ in this work, $c_a$ is the adiabatic sound speed of dark energy, and $\br$ is the energy density ratio of dark matter to dark energy, i.e., $\br =\rho_c/\rho_d$.

\section{The KiDS and Planck datasets}

The KiDS is optimally designed to measure shapes of galaxies with photometric redshifts so a study of weak lensing tomography can be performed. In this work we use angular correlation functions measured from KiDS's 450 degree square data to constrain our IDMDE models.

The angular shear correlation function $\xi_{\pm}^{ij}$ between redshifts $i$ and $j$ is given by the convergence power spectrum via
\begin{equation}
\label{eq.corP}
\xi_{\pm}^{ij}(\theta)=\frac{1}{2\pi}\int \text{d}llP_{\kappa}^{ij}J_{0,4}(l\theta),
\end{equation}
where $\theta$ is the angular position on the sky, $l$ is the angular wave number, and $J_{0,4}(l\theta)$ are the zeroth and the fourth order Bessel functions of the first kind for $\xi_+$ and $\xi_-$, respectively. According to the Limber
approximation, the convergence power spectrum $P_{\kappa}^{ij}$ can be written as
\begin{equation}
\label{eq.PP}
P_{\kappa}^{ij}=\int_0^{\chi_H}\text{d}\chi\frac{W_i(\chi)W_j(\chi)}{\chi^2}P_{\delta}\left(\frac{l}{\chi},\chi\right),
\end{equation}
where $\chi$ is comoving distance, $\chi_H$ is the comoving distance evaluated at an infinite redshift, $W_i(\chi)$ is the lensing weighting function corresponding to a redshift bin $i$,  and $P_{\delta}$ is a non-linear matter power spectrum derived from a IDMDE model. The lensing weighting function $W_i(\chi)$ is \cite{Hildebrandt2017}
\begin{equation}
\label{eq.qLCDM}
W_i(\chi)=\frac{3H_0^2\Omega_{m0}}{2c^2a(\chi)}\chi \int_{\chi}^{\chi_H}\text{d}\chi'n_i(\chi')\frac{\chi'-\chi}{\chi'},
\end{equation}
for the standard $\Lambda$CDM model. But this expression is not valid when there is interaction between dark sectors, since the relation of the standard dark matter density evolution $\Omega_c/\Omega_{c0}=H_0^2/(a^3H^2)$ fails for interacting dark energy models. Instead, equation (9) should be written in the general form \cite{Schafer2008}
\begin{equation}
\label{eq.q}
W_i(\chi)=\frac{3a(\chi)^2H(\chi)^2\Omega_m(\chi)}{2c^2}\chi\int_{\chi}^{\chi_H}\text{d}\chi'n_i(\chi')\frac{\chi'-\chi}{\chi'},
\end{equation}
where $\Omega_m=\rho_m/\rho_{\rm{crit}}$, the critical density $\rho_{\rm{crit}}=3H^2/(8\pi G)$, $c$ is the speed of light, $n_i$ is the galaxy redshift distribution in the bin $i$ and it is normalized as $\int_0^{\chi_H}n(\chi)\text{d}\chi=1$.

The KiDS-450 datasets consist of four tomographic redshift bins ($0.1<z<0.3$, $0.3<z<0.5$, $0.5<z<0.7$, $0.7<z<0.9$), and nine angular bins centered at $\theta$ = ($0.7134'$, $1.452'$, $2.956'$, $6.017'$, $12.25'$, $24.93'$, $50.75'$, $103.3'$, $210.3'$). For each tomographic redshift pair ($ij$), the angular ranges are limited to $\theta<72'$ for $\xi_+^{ij}$ and $\theta>4.2'$ for $\xi_-^{ij}$, so the last two bins and first three bins for $\xi_+^{ij}$ and $\xi_-^{ij}$ are masked out, respectively. Eventually there are $130$ angular band powers in this datasets which can be used to constrain the IDMDE models \cite{Joudaki2017}.

In addition to the KiDS-450 datasets, we also use the latest CMB power spectra from Planck 2015 data release~\cite{Adam2016} to derive constraints, which can be directly compared to the previous ones, for our IDMDE models~\cite{Costa2013,Costa2017}. Similar to ~\cite{Costa2017}, we take all the CMB temperature and polarization power spectra within $2<\ell<2000$ except the $BB$ power spectrum for which only the low-$\ell$, i.e. $2<\ell<30$, measurement is made available.

To sample parameter space of our cosmological models, we carry out a series of Markov Chain Monte Carlo (MCMC) runs using the modified CosmoMC code package \footnotemark[1]\footnotetext[1]{https://github.com/sjoudaki/kids450} that has already integrated the weak lensing module as described in~\cite{Joudaki2017}. We assume a flat universe with no running of the spectral index. We fix the effective number of neutrino species to $N_{\rm{eff}} = 3.046$, the sum of neutrino masses to $\Sigma m_{\nu}= 0.06\,\rm{eV}$, and the primordial helium fraction $Y_{\rm p}=0.25$. For the MCMC runs, a convergence criterion is set to $R-1 = 0.02$ where $R$ is the Gelman--Rubin threshold~\cite{Andrew1992}.

Before we launch the MCMC runs, priors on cosmological parameters are chosen and listed in Table \ref{tab.prior}. These priors are chosen the same as the ones in~\cite{Costa2017} because we want to validate that our analysis with the same model, so that comparisons can be carried out by constraining the IDMDE models using the Planck data available in \cite{Costa2017} and the weak lensing data, and examining the concordance problem with these two different datasets.

\begin{table}[ht]
\caption{Priors on cosmological parameters for phenomenological IDMDE models \label{tab.prior}}
\begin{tabular}{c|c|c|c|c}
\hline
Parameter & \multicolumn{4}{|c}{Prior} \\
\hline
$\Omega_bh^2$ & \multicolumn{4}{|c}{[0.005,0.1]}\\
\hline
$\Omega_ch^2$ & \multicolumn{4}{|c}{[0.001,0.99]}\\
\hline
$100\theta$ & \multicolumn{4}{|c}{[0.5,10]}\\
\hline
$\tau$ & \multicolumn{4}{|c}{[0.01,0.8]}\\
\hline
$n_s$ & \multicolumn{4}{|c}{[0.9,1.1]}\\
\hline
$\text{log}(10^{10}A_s)$ & \multicolumn{4}{|c}{[2.7,4]}\\
\hline
$h$ & \multicolumn{4}{|c}{[0.4,1]}\\
\hline
\hline
Model & I & II & III & IV\\
\hline
$w$ & [-1,-0.3] & [-3,-1] & [-3,-1] & [-3,-1]\\
\hline
$\lambda_i$ & [-0.4,0] & [0,0.4] & [0.0.01] & [0,0.01]\\
\hline
\end{tabular}
\end{table}

\section{Results}

\begin{figure}[ht]
\centering
\includegraphics[scale=0.6]{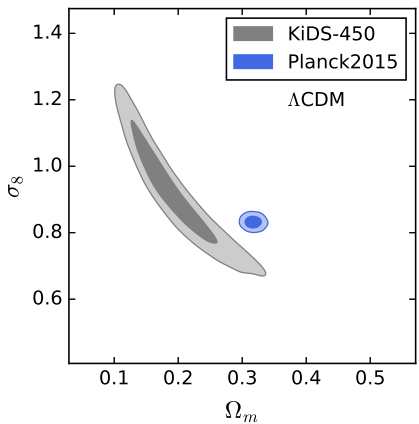}
\caption{Marginalized posterior distribution function in the  $\sigma_8-\Omega_m$ plane for the $\Lambda$CDM model. $68\%$ and 95\% confidence levels are shown as inner and outer regions. }
\label{fig.LCDMomsi8}
\end{figure}

Before investigating the IDMDE models, we first compare the theoretical predictions of the weak lensing correlation functions from the standard $\Lambda$CDM model to both KiDS and Planck measurements. In Fig. \ref{fig.LCDMomsi8} we show parameter constraints for KiDS in the $\sigma_8\mbox{--}\Omega_m$ plane using priors in Table \ref{tab.prior}, where the amplitude of scalar perturbation $A_s$ and the scalar spectral index $n_s$ are limited to much narrower parameter space, as compared to those in \cite{Joudaki2017}. These priors shrink the parameter contours and thereby increase the tension between KiDS weak lensing and Planck CMB temperature measurements for the $\Lambda$CDM model as Fig. \ref{fig.LCDMomsi8} clearly shows. The main reason for choosing the same priors in Table II, as Table 3 in \cite{Costa2017} instead of those exactly from \cite{Joudaki2017}, is to compare our model fitting to KiDS datasets with that to Planck datasets which was already available in \cite{Costa2017}, especially in the following discussions for the IDMDE models.

In this section, we will investigate the IDMDE models and examine how these models are favored by both datasets and to what extent they will alleviate the tension.

\subsection{Model selection}

We now use the IDMDE models to predict the CMB power spectra and weak lensing shear correlation functions and compare them to the Planck and the KiDS-450 datasets, and the two sets of constraints that are derived from the IDMDE and the standard $\Lambda$CDM models will be passed along the Deviance Information Criterion (DIC) \cite{Spiegelhalter2002,Joudaki2016} to determine which model is more favored. The DIC is composed of the sum of goodness of fit of a given model and its Bayesian complexity, and it is defined as
\begin{equation}
\label{eq.DIC}
{\rm DIC}=\chi_{\rm eff}^2(\bm{\hat\theta})+2p_{\text{D}},
\end{equation}
where $\chi_{\rm{eff}}^2(\bm{\hat\theta})=-2\text{ln}\mathcal{L}_{\rm max}$ is the best-fit effective $\chi^2$ and $\bm{\hat\theta}$ is the parameter vector at the maximum likelihood point. The generic form of $\chi^2$ is expressed as $\Delta\bm{\xi}^T \text{\bf Cov}^{-1}\Delta\bm{\xi}$, where $\Delta\bm{\xi}=\bm{\xi}_{\rm obs}-\bm{\xi}_{\rm theory}$, $\bm{\xi}_{\rm obs}$ is a 130-element vector of the measured weak lensing correlation functions from the KiDS project, $\bm{\xi}_{\rm theory}$ is the theoretical prediction from equation (\ref{eq.corP}), and {\bf Cov}, which is a 130$\times$130 covariance matrix at each angular scale, is also provided by the KiDS data release.

The second term in equation (\ref{eq.DIC}) is the Bayesian complexity expressed as $p_{\text{D}}=\langle\chi_{\rm eff}^2(\bm{\theta})\rangle-\chi_{\rm eff}^2(\bm{\hat\theta})$, where $\langle\chi_{\rm eff}^2(\bm{\theta})\rangle$ represents the mean $\chi^2$ averaged over the posterior distribution. We define the differences in DIC as
\begin{equation}
\Delta{\rm DIC}={\rm DIC}({\rm IDMDE})-{\rm DIC}(\Lambda{\rm CDM}),
\end{equation} and the condition of $\Delta{\rm DIC} < -5$ indicates a moderate preference in favor of the IDMDE model while the condition $\Delta{\rm DIC} \sim 0$ means that one model is not favored over the other \cite{Joudaki2017}.

In Table \ref{tab.changes}, we calculate changes relative to the reference $\Lambda$CDM model in both $\chi_{\rm eff}^2(\bm{\hat\theta})$ and DIC for four IDMDE models, where the negative values of DIC indicate preference in favor of the IDMDE models. In all of the IDMDE models, we find that either KiDS or Planck on its own does not show any preference for the IDMDE models because $|\Delta\text{DIC}|$ is always smaller than $5$. But the combination of KiDS and Planck datasets does favor the IDMDE models with strong negative $\Delta$DIC values.

\begin{table}[ht]
\caption{$\Delta\chi_{\rm eff}^2(\bm{\hat\theta})$ and $\Delta$DIC between the IDMDE models and the reference $\Lambda$CDM model\label{tab.changes}. $\Delta\chi^2_{\rm eff}$ is the minimum $\chi^2$ difference between the model and $\Lambda$CDM. }
\begin{tabular}{c|lcc}
\hline
\ \ Model \ \ & \ \ Data & \ \ \ \ $\Delta\chi_{\rm eff}^2$ \ \ \ \ &\ \ \ \  $\Delta$DIC\ \ \ \  \\
\hline
\multirow{3}{*}{I}
& \ \ KiDS & $2.016$ & $6.02$ \\
& \ \ Planck & $1.368$ & $3.856$ \\
& \ \ KiDS+Planck & $-19.094$ & $-24.299$ \\
\hline
\multirow{3}{*}{II}
& \ \ KiDS & $-0.813$ & $0.578$ \\
& \ \ Planck & $-2.676$ & $-1.735$ \\
& \ \ KiDS+Planck & $-5.762$ & $-9.380$ \\
\hline
\multirow{3}{*}{III}
& \ \ KiDS & $-1.174$ & $0.480$ \\
& \ \ Planck & $-1.292$ & $-0.611$ \\
& \ \ KiDS+Planck & $-5.460$ & $-10.808$ \\
\hline
\multirow{3}{*}{IV}
& \ \ KiDS & $-0.710$ & $0.511$ \\
& \ \ Planck & $-1.550$ & $-0.816$ \\
& \ \ KiDS+Planck & $-3.986$ & $-12.385$ \\
\hline
\end{tabular}
\end{table}

In Fig. (\ref{fig.omsi8}), we show the parameter constraints in the $\sigma_8-\Omega_m$ plane for all of the IDMDE models. The KiDS constraints are in gray and the Planck constraints are in blue. Different from the constraints of the $\Lambda$CDM model in Fig. (\ref{fig.LCDMomsi8}), Fig. (\ref{fig.omsi8}) shows that the KiDS and Planck constraints of the IDMDE models start to overlap with each other.

In order to quantify at what level the tension between KiDS and Planck has been reduced by the IDMDE models, a tension parameter $T$ is adopted and it is defined as~\cite{Joudaki2017}
\begin{equation}
\label{eq.TS8}
T(S_8)=\frac{|\langle S_8^{\text{K}}\rangle-\langle S_8^{\text{P}}\rangle|}{\sqrt{\sigma^2(S_8^{\text{K}})+\sigma^2(S_8^{\text{P}})}}.
\end{equation}
Here $S_8=\sigma_8\sqrt{\Omega_m}$, $\langle S_8\rangle$ is the mean value over the posterior distribution, $\sigma$ is the standard deviation, and the superscripts K and P correspond to KiDS and Planck, respectively.

\begin{figure}[ht]
\centering
\includegraphics[scale=0.6]{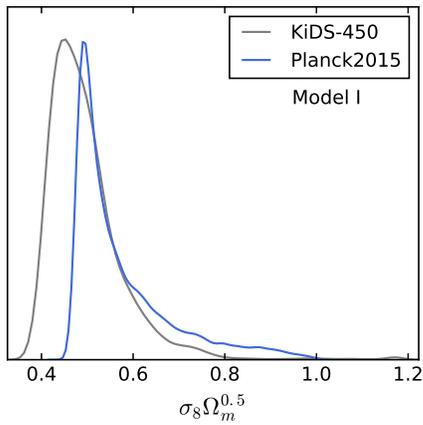}
\caption{1D distributions of $S_8$ for KiDS and Planck.}
\label{fig.s8_I}
\end{figure}

In Table \ref{tab.T}, we list all the values of $T(S_8)$ for $\Lambda$CDM and the IDMDE models. It is quite obvious that the IDMDE models can reduce discordance significance to $\sim1\sigma$ level, substantially relieving the $2\sigma$-level tension inferred from the standard $\Lambda$CDM model. Among all the IDMDE models, the $T$-parameter values can further indicate that Model I has the least discordance between two datasets while other IDMDE models have roughly the same level of discordance. The overlapped contours are shown in Fig. (\ref{fig.omsi8}).

In Model I, the blue contours from Planck show significantly different directions compared to that in the $\Lambda$CDM model because the posterior distribution functions are considerably skewed compared with other IDMDE models. And this is verified in Fig. (\ref{fig.s8_I}), where probability distribution of the parameter $S_8$  is not symmetric. So we should note that the $T$-parameter defined in Eq. (\ref{eq.TS8}) may not be a very appropriate method for quantifying the discordance of Model I.

From both the DIC and the $T$-parameter tests, we find that the IDMDE models can effectively alleviate the tensions between KiDS and Planck datasets, thereby they are more favored than the  $\Lambda$CDM model by the data.

\begin{table}[ht]
\caption{The values of $T(S_8)$ for the $\Lambda$CDM and four interacting dark energy models shown in Table \ref{tab.model}. \label{tab.T}}
\begin{tabular}{c|ccccc}
\hline
\ \ \ \ Model\ \ \ \  & $\Lambda$CDM & I & II & III & IV \\
\hline
$T(S_8)$ &\ \  $2.11\sigma$\ \  &\ \  $0.44\sigma$\ \  &\ \  $1.08\sigma$\ \  &\ \  $1.18\sigma$\ \  &\ \  $1.24\sigma$\ \  \\
\hline
\end{tabular}
\end{table}

\begin{figure*}
 \includegraphics[scale=0.45]{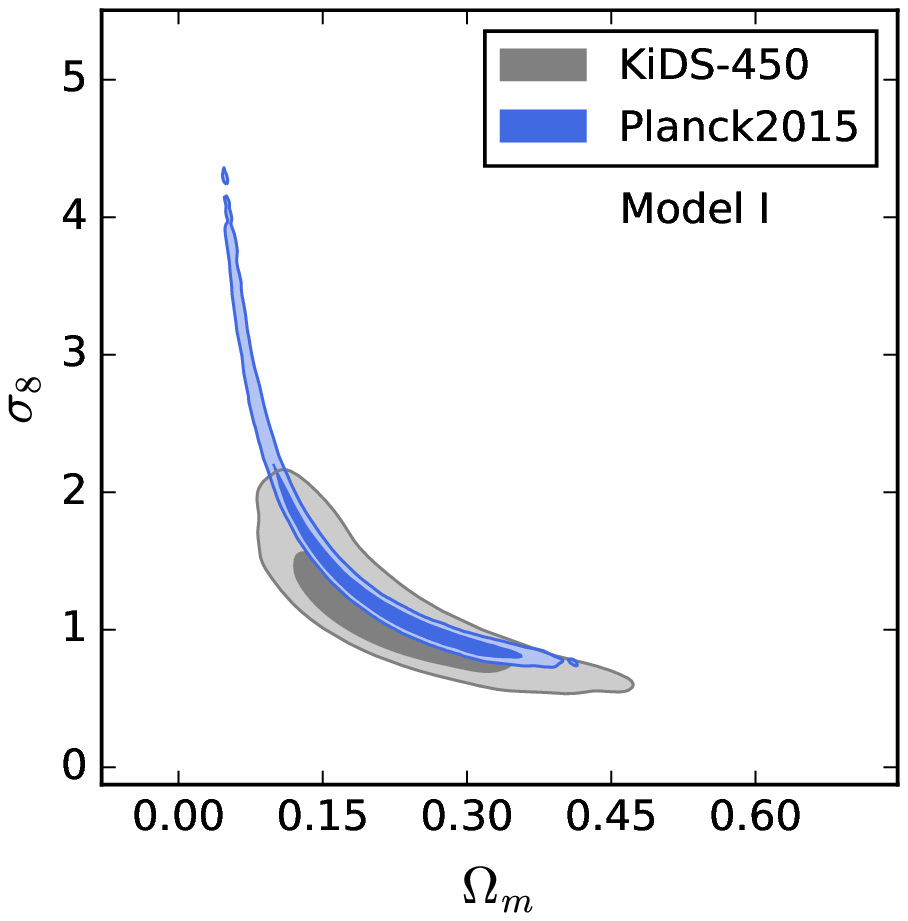}
 \includegraphics[scale=0.45]{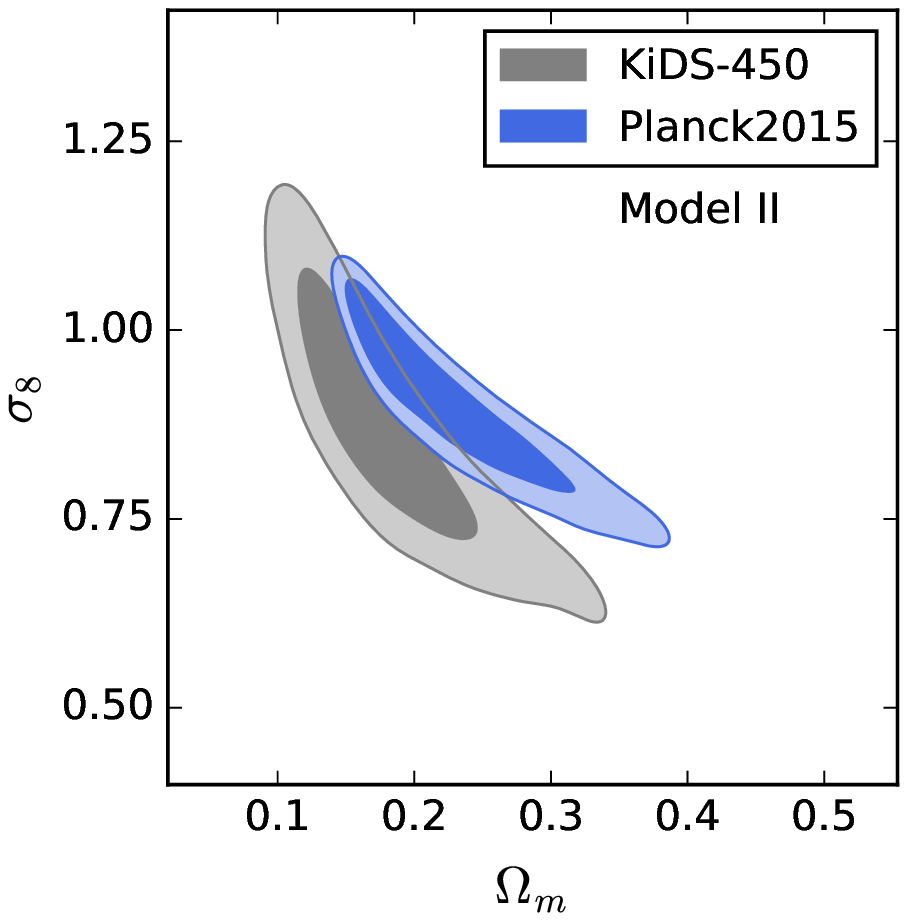}
 \includegraphics[scale=0.45]{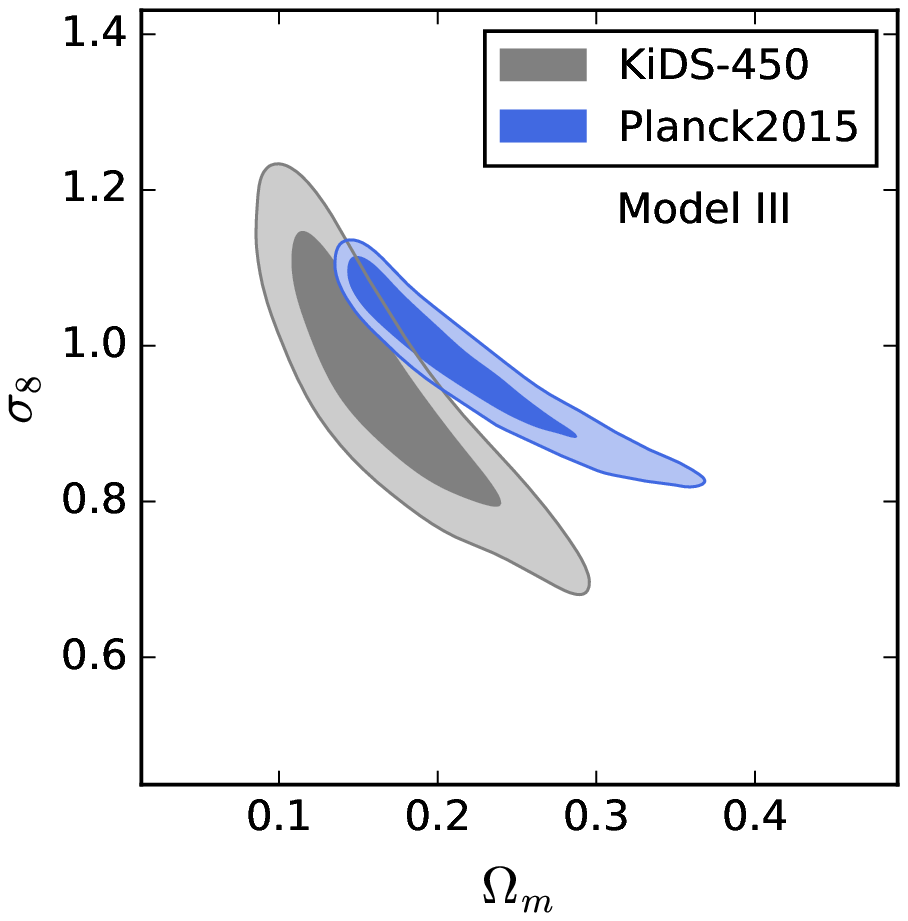}
 \includegraphics[scale=0.45]{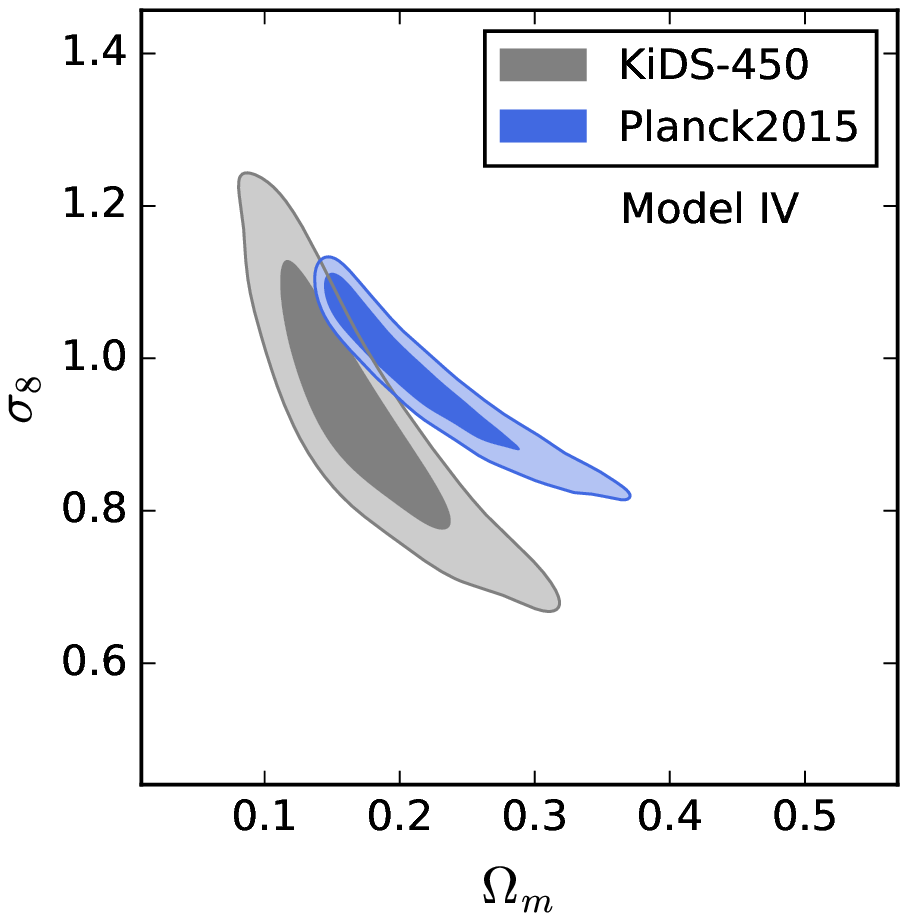}
\caption{Marginalized confidence contours in the  $\sigma_8-\Omega_m$ plane for four IDMDE models. $68\%$ and 95\% confidence levels are shown as inner and outer regions. }
\label{fig.omsi8}
\end{figure*}

\subsection{Model constraints}

\begin{figure*}
\centering
\subfloat[Model I]{%
 \includegraphics[scale=0.4]{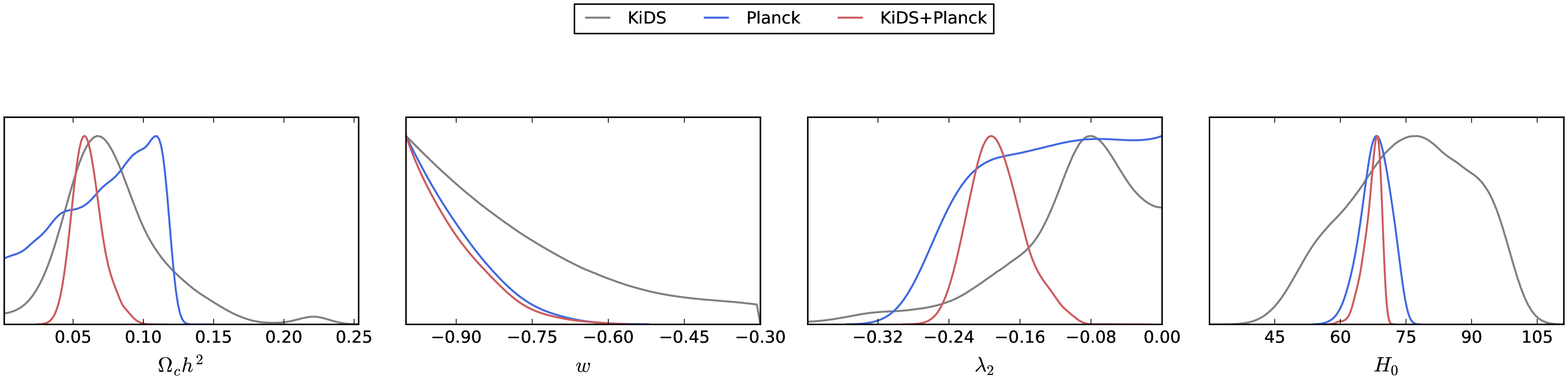}}\hfill
\subfloat[Model II]{%
 \includegraphics[scale=0.4]{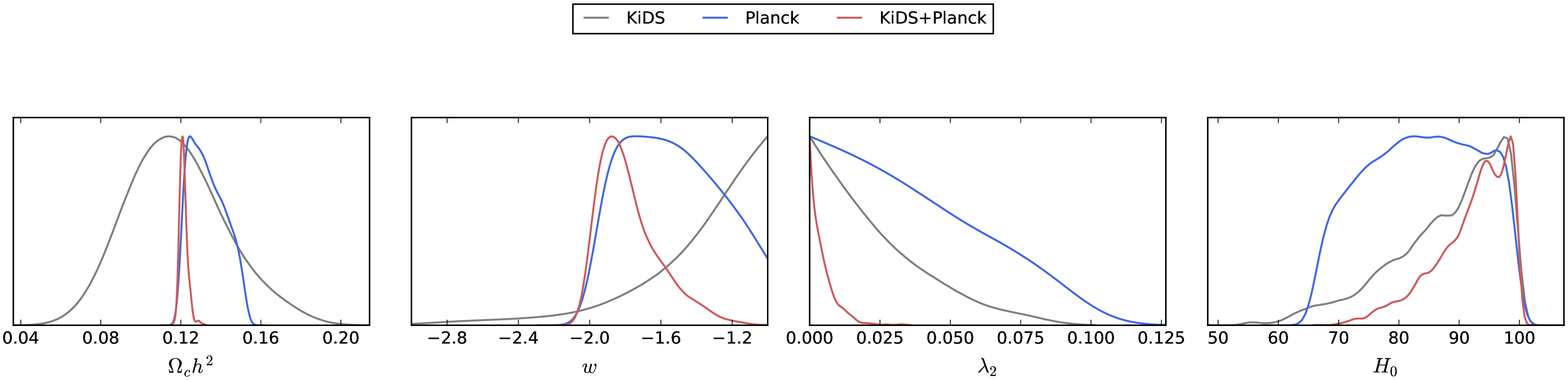}}\hfill
\subfloat[Model III]{%
 \includegraphics[scale=0.4]{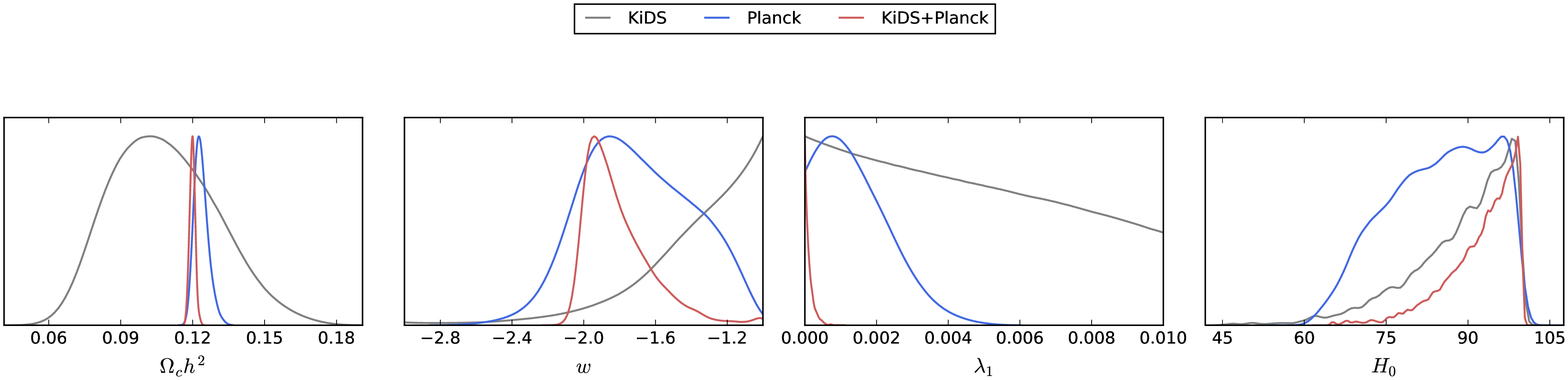}}\hfill
\subfloat[Model IV]{%
 \includegraphics[scale=0.4]{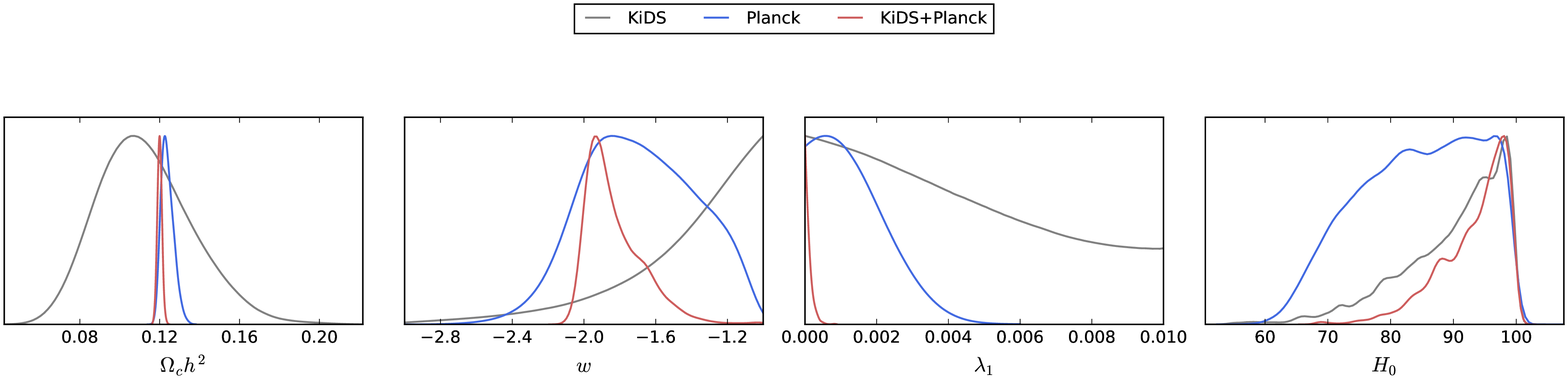}}
\caption{1D likelihood functions for selected parameters}
\label{fig.1D}
\end{figure*}

\begin{figure*}
\centering
\subfloat[Model I]{%
 \includegraphics[scale=0.5]{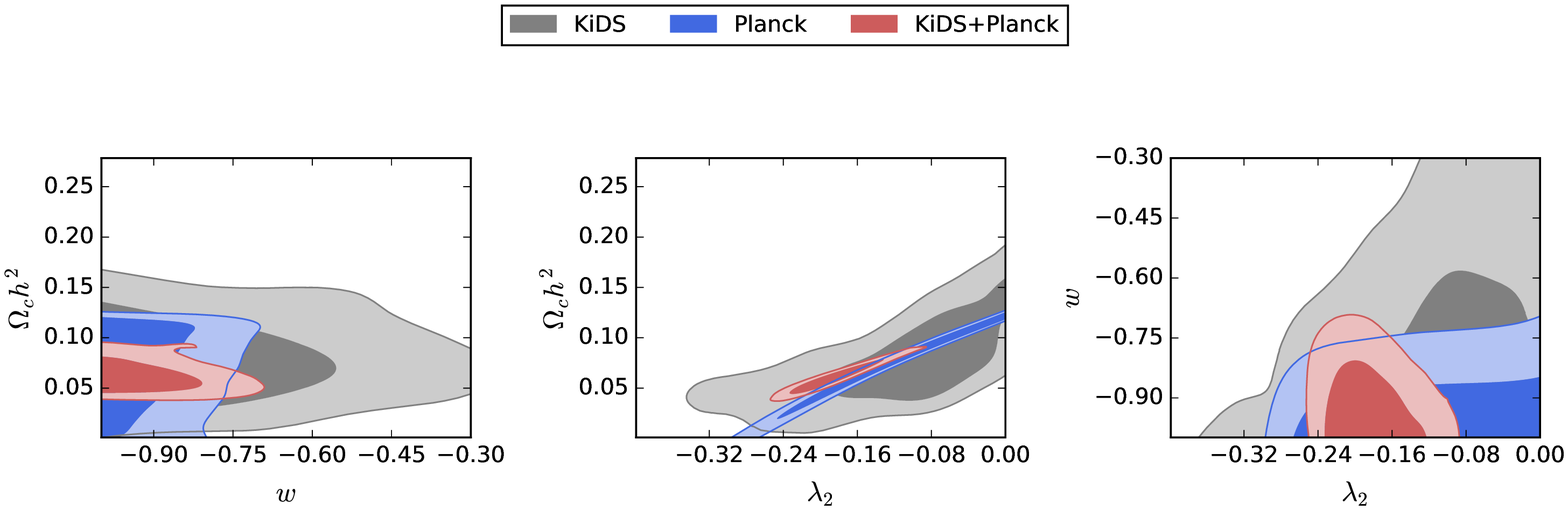}}\hfill
\subfloat[Model II]{%
 \includegraphics[scale=0.5]{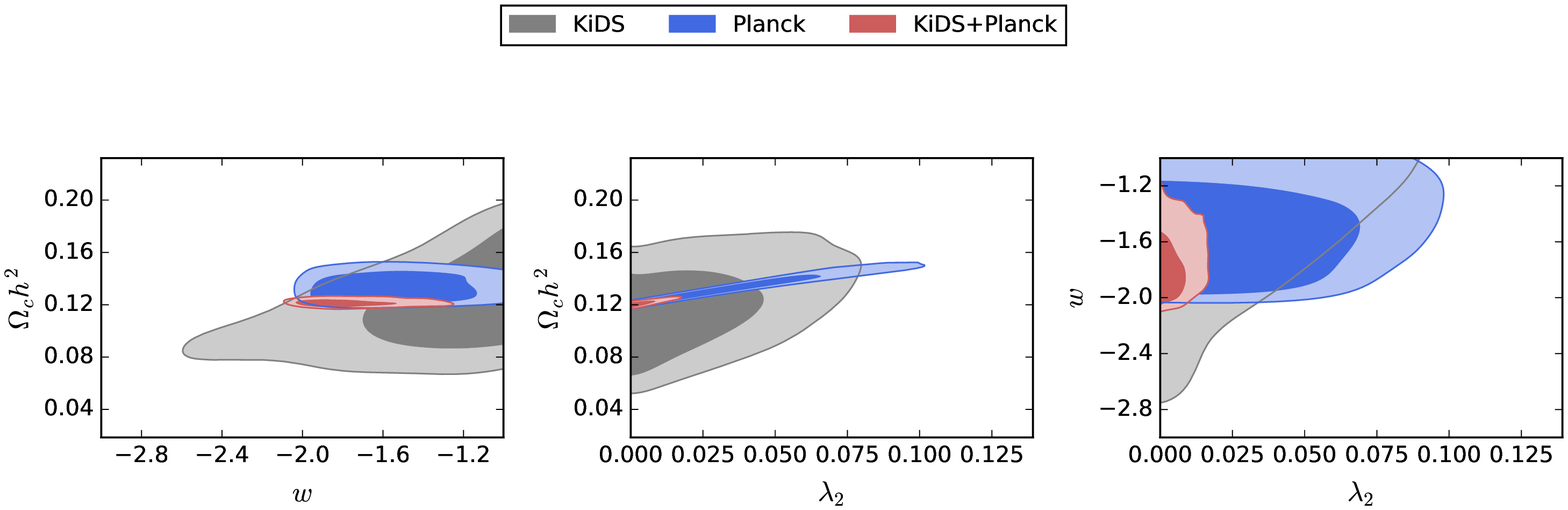}}\hfill
\subfloat[Model III]{%
 \includegraphics[scale=0.5]{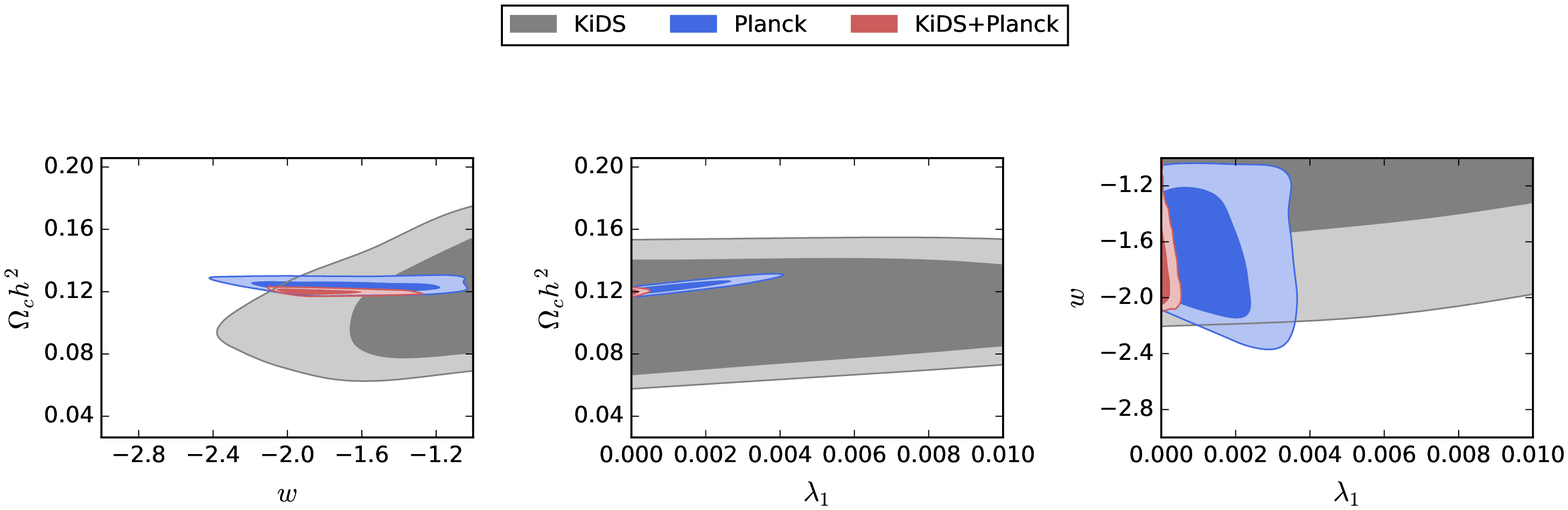}}\hfill
\subfloat[Model IV]{%
 \includegraphics[scale=0.5]{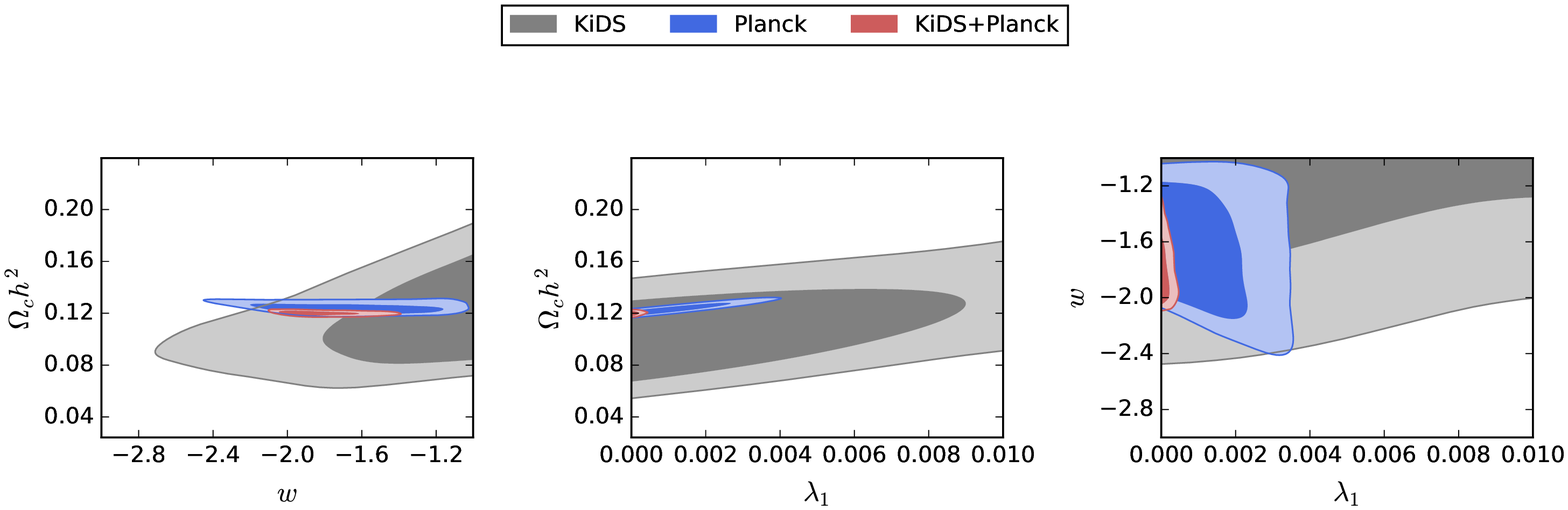}}
\caption{2D posterior contours for selected parameters}
\label{fig.2D}
\end{figure*}

In this section, we present all the results from the MCMC runs. We derive constrains for the IDMDE models from the  KiDS-450 datasets and compare them to the previous limits set by Planck data \cite{Costa2017}. We perform the joint analysis with both the KiDS and Planck data and list all the best-fit values and 68\% confidence regions in Tables \ref{tab.I}, \ref{tab.II}, \ref{tab.III} and \ref{tab.IV}. The 1D marginalized posterior distribution functions (PDFs) are shown for the parameters $\Omega_ch^2$, $w$, $\lambda_{i}$, and $H_0$ in Fig. (\ref{fig.1D}) and the 2D functions in Fig. (\ref{fig.2D}).

(1) Model I: the 1D PDFs are shown for the parameters $\Omega_ch^2$, $w$, $\lambda_{2}$, and $H_0$ in Fig. (\ref{fig.1D}\textcolor{blue}{a}) and the 2D PDFs in Fig. (\ref{fig.2D}\textcolor{blue}{a}). Due to large band power uncertainties, we find that the constraints from the KiDS data alone have wider $1\sigma$ and $2\sigma$ contours compared to Planck for most of the parameters. However, we find that the $68\%$ confidence regions of $\Omega_ch^2$, $\lambda_2$, $\Omega_{d}$ and $\Omega_{m}$ from KiDS data (in Table \ref{tab.I}) are smaller than constraints from the Planck data and this indicates that the weak lensing measurements are more sensitive to the interaction and the energy densities of dark sectors compared to the cosmic microwave background measurements. Among all the constraints from both the KiDS and Planck datasets, Model I is of particular interest in that KiDS data alone can give a negative coupling parameter $\lambda_2$, which is in agreement with the Planck constraint \cite{Costa2017}, and the sign of the coupling does not change even if the KiDS and Planck datasets are combined. The negative coupling indicates that the energy flows from dark matter to dark energy, unfortunately, the IDMDE models with negative couplings are not favored to alleviate  the coincidence problem, see review and references therein for detailed explanations \cite{Wang2016}. Therefore, Model I can achieve a concordance between observational datasets at the cost of intensifying the coincidence problem.

(2) Models II, III and IV: the 1D and 2D marginalized PDFs are shown in Figs. (\ref{fig.1D}\textcolor{blue}{b}, \ref{fig.1D}\textcolor{blue}{c}, \ref{fig.1D}\textcolor{blue}{d}) and Figs. (\ref{fig.2D}\textcolor{blue}{b}, \ref{fig.2D}\textcolor{blue}{c}, \ref{fig.2D}\textcolor{blue}{d}), respectively. Similar to Model I, the Planck measurements always have tighter constraints on most of the parameters as compared to the KiDS measurements. The positive sign of the constrained parameter $\lambda_2$ indicates that there is energy transferred from dark energy to dark matter so the duration for the energy densities of dark matter and dark energy being comparable is much longer than $\Lambda$CDM, alleviating the coincidence problem~\cite{Costa2013}.

The KiDS-only constraint on $\sigma_8$ for the standard model is $\sigma_8=0.849_{-0.204}^{+0.120}$ \cite{Hildebrandt2017}, and its constraint on $H_0$ of the standard model is also found reasonable therein. Now with the extra degree of freedom introduced in the IDMDE models, Table \ref{tab.I} shows that the KiDS-only constraint on $\sigma_8$ in Model I is quite large, inconsistent with other cosmological observations, and so are the values of the Hubble constant $H_0$ that are shown in Tables \ref{tab.II}, \ref{tab.III} and \ref{tab.IV}, corresponding to the three IDMDE models. Despite high values, these results are still consistent with the previous Planck-only constraints \cite{Costa2017}. Also, these high values, which could be considered as unreasonable constraints, can be easiliy reduced by adding more observational datasets such as BAO and SNIa, as done in  \cite{Costa2017}. Similarly, the high values derived from the KiDS-only measurements could be reduced as well, by adding those observations.  However, in the scope of this work, we exclusively focus on the discordance problem found in the standard $\Lambda$CDM model between weak lensing and Planck datasets, not the most stringent limits from multiple observational datasets. Specifically, our motivation is to investigate if the discordance can be alleviated by the IDMDE models. From the analysis in this work, it is proven that the tension between Planck and KiDS datasets is indeed alleviated.

\section{Conclusions}

In this work we have obtained the observational constraints on the IDMDE models using both weak gravitational lensing data from the KiDS and the CMB data from Planck. We have examined how the IDMDE models are favored by data as compared to the standard $\Lambda$CDM and to what extent the IDMDE models can relieve the tensions between KiDS and Planck CMB datasets. Employing the DIC and $T$-parameter diagnostics, we find with great interest that a desired concordance between KiDS and Planck datasets can be achieved by the IDMDE models which are more preferred than the standard $\Lambda$CDM model.

We obtain parameter constraints on the IDMDE models from KiDS measurements, Planck measurements and the combination. Due to large band-power uncertainties in the weak lensing measurements, the KiDS data alone always have less constraining power than the Planck datasets for most of the parameters but a joint fit to both datasets can always improve the constraints significantly. For model I, the sign of the interaction parameter being negative indicates that the energy flows from dark matter to dark energy. And this scenario is not favored to the alleviation of the coincidence problem. While the other three models favor a scenario with energy flowing from dark energy to dark matter, which thus can alleviate the coincidence problem.

\begin{acknowledgments}
We thank Jun Zhang and Zuhui Fan for helpful comments and suggestions. This work was partially supported by National Basic Research Program of China (973 Program 2013CB834900) and National Natural Science Foundation of China.
\end{acknowledgments}

\begin{table*}
\caption{Best fit values and $68\%$ confidence levels for parameters in Model I. \label{tab.I}}
\begin{tabular}{c|cc|cc|cc}
\hline
 & \multicolumn{2}{|c|}{KiDS} & \multicolumn{2}{|c|}{Planck} & \multicolumn{2}{|c}{KiDS+Planck} \\
\hline
Parameter & Best fit & $68\%$ limits & Best fit & $68\%$ limits & Best fit & $68\%$ limits\\
\hline
$\Omega_bh^2$ & $0.01380$ & $0.04324_{-0.03824}^{+0.01186}$ & $0.02231$ & $0.0222_{-0.00016}^{+0.00016}$ & $0.02119$ & $0.02121_{-0.000174}^{+0.000152}$ \\
$\Omega_ch^2$ & $0.09104$ & $0.08022_{-0.04046}^{+0.02162}$ & $0.04788$ & $0.07131_{-0.024}^{+0.0472}$ & $0.06127$ & $0.06287_{-0.01427}^{+0.00744}$ \\
$100\theta$ & $1.108$ & $1.033_{-0.06933}^{+0.05013}$ & $1.045$ & $1.044_{-0.00329}^{+0.0015}$ & $1.044$ & $1.044_{-0.000665}^{+0.001104}$ \\
$\tau$ & $0.4419$ & $0.4122_{-0.4022}^{+0.1533}$ & $0.08204$ & $0.08063_{-0.0169}^{+0.0171}$ & $0.04309$ & $0.04377_{-0.02402}^{+0.0164}$ \\
$\text{ln}(10^{10}A_s)$ & $3.636$ & $3.249_{-0.5494}^{+0.1710}$ & $3.102$ & $3.097_{-0.0329}^{+0.0328}$ & $3.020$ & $3.021_{-0.0472}^{+0.0349}$ \\
$n_s$ & $0.9435$ & $0.9931_{-0.09315}^{+0.1068}$ & $0.9639$ & $0.9633_{-0.00514}^{+0.00472}$ & $0.9453$ & $0.9495_{-0.00511}^{+0.00465}$ \\
$w$ & $-0.8682$ & $-0.7811_{-0.2178}^{+0.05257}$ & $-0.9765$ & $-0.9031_{-0.0959}^{+0.023}$ & $-0.9554$ & $-0.9079_{-0.082}^{+0.0217}$ \\
$\lambda_2$ & $-0.01512$ & $>-0.1237$ & $-0.1831$ & $>-0.1745$ & $-0.1871$ & $-0.1780_{-0.0456}^{+0.0215}$ \\
\hline
$H_0$ & $84.35$ & $76.10_{-11.95}^{+17.62}$ & $72.36$ & $68.1_{-3.2}^{+2.99}$ & $68.46$ & $67.27_{-1.16}^{+2.44}$ \\
$\Omega_d$ & $0.8517$ & $0.7792_{-0.05074}^{+0.09475}$ & $0.8647$ & $0.7899_{-0.106}^{+0.0932}$ & $0.8227$ & $0.8119_{-0.0163}^{+0.0484}$ \\
$\Omega_m$ & $0.1482$ & $0.2207_{-0.09475}^{+0.05074}$ & $0.1353$ & $0.2101_{-0.0926}^{+0.106}$ & $0.1773$ & $0.1881_{-0.0484}^{+0.0163}$ \\
$\sigma_8$ & $1.107$ & $1.131_{-0.3978}^{+0.1883}$ & $1.622$ & $1.438_{-0.789}^{+0.143}$ & $1.385$ & $1.359_{-0.191}^{+0.235}$ \\
Age/Gyr & $12.74$ & $12.52_{-2.311}^{+1.531}$ & $13.71$ & $13.81_{-0.0916}^{+0.058}$ & $13.86$ & $13.91_{-0.0496}^{+0.0288}$ \\
\hline
\end{tabular}
\end{table*}

\begin{table*}
\caption{Best fit values and $68\%$ confidence levels for parameters in Model II. \label{tab.II}}
\begin{tabular}{c|cc|cc|cc}
\hline
 & \multicolumn{2}{|c|}{KiDS} & \multicolumn{2}{|c|}{Planck} & \multicolumn{2}{|c}{KiDS+Planck} \\
\hline
Parameter & Best fit & $68\%$ limits & Best fit & $68\%$ limits & Best fit & $68\%$ limits\\
\hline
$\Omega_bh^2$ & $0.01020$ & $0.02355_{-0.01855}^{+0.003478}$ & $0.02232$ & $0.02225_{-0.000161}^{+0.000162}$ & $0.02137$ & $0.02149_{-0.000141}^{+0.000122}$ \\
$\Omega_ch^2$ & $0.1063$ & $0.1176_{-0.02790}^{+0.02170}$ & $0.1314$ & $0.1334_{-0.0125}^{+0.00692}$ & $0.12033$ & $0.12136_{-0.00226}^{+0.00134}$ \\
$100\theta$ & $1.061$ & $1.034_{-0.05481}^{+0.05504}$ & $1.04$ & $1.04_{-0.000562}^{+0.000651}$ & $1.040$ & $1.040_{-0.000309}^{+0.000321}$ \\
$\tau$ & $0.2427$ & $0.4004_{-0.3876}^{+0.3974}$ & $0.07543$ & $0.07653_{-0.0174}^{+0.0177}$ & $0.01224$ & $0.02201_{-0.01201}^{+0.00306}$ \\
$\text{ln}(10^{10}A_s)$ & $3.466$ & $3.151_{-0.4516}^{+0.1323}$ & $3.082$ & $3.088_{-0.0337}^{+0.0342}$ & $2.951$ & $2.967_{-0.0245}^{+0.0109}$ \\
$n_s$ & $1.039$ & $1.015_{-0.02851}^{+0.08414}$ & $0.9657$ & $0.9638_{-0.00475}^{+0.00477}$ & $0.9547$ & $0.9575_{-0.00408}^{+0.00406}$ \\
$w$ & $-1.370$ & $-1.397_{-0.06139}^{+0.06139}$ & $-1.872$ & $-1.55_{-0.358}^{+0.235}$ & $-1.977$ & $-1.780_{-0.215}^{+0.089}$ \\
$\lambda_2$ & $0.02176$ & $<0.00294$ & $0.02931$ & $<0.05044$ & $0.00008$ & $<0.00564$ \\
\hline
$H_0$ & $99.52$ & $88.13_{-3.988}^{+11.64}$ & $96.2$ & $83.88_{-7.86}^{+13.3}$ & $98.3$ & $92.04_{-2.36}^{+7.81}$ \\
$\Omega_d$ & $0.8816$ & $0.8135_{-0.03261}^{+0.06042}$ & $0.8331$ & $0.7688_{-0.0353}^{+0.0778}$ & $0.8527$ & $0.828_{-0.0067}^{+0.0305}$ \\
$\Omega_m$ & $0.1183$ & $0.1864_{-0.06042}^{+0.03261}$ & $0.1669$ & $0.2312_{-0.0778}^{+0.0353}$ & $0.1473$ & $0.172_{-0.0305}^{+0.0067}$ \\
$\sigma_8$ & $1.046$ & $0.8548_{-0.1311}^{+0.1091}$ & $0.9852$ & $0.9016_{-0.094}^{+0.0945}$ & $1.0283$ & $0.9725_{-0.0382}^{+0.061}$ \\
Age/Gyr & $13.62$ & $13.58_{-1.845}^{+1.181}$ & $13.46$ & $13.59_{-0.143}^{+0.0708}$ & $13.55$ & $13.38_{-0.064}^{+0.028}$ \\
\hline
\end{tabular}
\end{table*}

\begin{table*}
\caption{Best fit values and $68\%$ confidence levels for parameters in Model III. \label{tab.III}}
\begin{tabular}{c|cc|cc|cc}
\hline
 & \multicolumn{2}{|c|}{KiDS} & \multicolumn{2}{|c|}{Planck} & \multicolumn{2}{|c}{KiDS+Planck} \\
\hline
Parameter & Best fit & $68\%$ limits & Best fit & $68\%$ limits & Best fit & $68\%$ limits\\
\hline
$\Omega_bh^2$ & $0.006800$ & $0.02134_{-0.01634}^{+0.002548}$ & $0.0223$ & $0.02235_{-0.00017}^{+0.00017}$ & $0.0214$ & $0.02148_{-0.000137}^{+0.000133}$ \\
$\Omega_ch^2$ & $0.1007$ & $0.1078_{-0.02530}^{+0.01939}$ & $0.1198$ & $0.1236_{-0.00353}^{+0.00235}$ & $0.1207$ & $0.0120_{-0.00114}^{+0.0012}$ \\
$100\theta$ & $1.094$ & $1.050_{-0.05642}^{+0.05750}$ & $1.041$ & $1.041_{-0.000374}^{+0.000377}$ & $1.041$ & $1.040_{-0.000306}^{+0.000306}$ \\
$\tau$ & $0.03344$ & $0.4049_{-0.3949}^{+0.3950}$ & $0.07784$ & $0.07051_{-0.0179}^{+0.0182}$ & $0.0149$ & $0.0218_{-0.011}^{+0.0021}$ \\
$\text{ln}(10^{10}A_s)$ & $3.565$ & $3.283_{-0.5839}^{+0.1875}$ & $3.087$ & $3.074_{-0.0355}^{+0.0357}$ & $2.956$ & $2.966_{-0.0233}^{+0.0103}$ \\
$n_s$ & $0.9392$ & $1.018_{-0.02662}^{+0.08126}$ & $0.9649$ & $0.9608_{-0.00503}^{+0.00508}$ & $0.9571$ & $0.9568_{-0.00406}^{+0.004}$ \\
$w$ & $-1.228$ & $-1.393_{-0.07466}^{+0.3929}$ & $-1.701$ & $-1.702_{-0.364}^{+0.298}$ & $-1.98$ & $-1.807_{-0.223}^{+0.0765}$ \\
$\lambda_1$ & $0.002013$ & $<0.006058$ & $0.0004372$ & $<0.001831$ & $0.00005$ & $<0.00013$ \\
\hline
$H_0$ & $99.28$ & $88.71_{-3.237}^{+11.16}$ & $89.51$ & $84.91_{-4.8}^{+15.1}$ & $98.1$ & $92.45_{-1.9}^{+6.55}$ \\
$\Omega_d$ & $0.8902$ & $0.8308_{-0.02877}^{+0.05582}$ & $0.8218$ & $0.788_{-0.0268}^{+0.0686}$ & $0.8516$ & $0.8308_{-0.0044}^{+0.0275}$ \\
$\Omega_m$ & $0.1098$ & $0.1691_{-0.05582}^{+0.02877}$ & $0.1782$ & $0.212_{-0.0686}^{+0.0268}$ & $0.1484$ & $0.1692_{-0.0275}^{+0.0044}$ \\
$\sigma_8$ & $1.201$ & $0.9538_{-0.1274}^{+0.1272}$ & $1.016$ & $0.9885_{-0.061}^{+0.102}$ & $1.034$ & $0.9905_{-0.0211}^{+0.0593}$ \\
Age/Gyr & $1.312$ & $13.64_{-1.823}^{+1.177}$ & $13.55$ & $13.71_{-0.18}^{+0.102}$ & $13.55$ & $13.60_{-0.065}^{+0.0941}$ \\
\hline
\end{tabular}
\end{table*}

\begin{table*}
\caption{Best fit values and $68\%$ confidence levels for parameters in Model IV. \label{tab.IV}}
\begin{tabular}{c|cc|cc|cc}
\hline
 & \multicolumn{2}{|c|}{KiDS} & \multicolumn{2}{|c|}{Planck} & \multicolumn{2}{|c}{KiDS+Planck} \\
\hline
Parameter & Best fit & $68\%$ limits & Best fit & $68\%$ limits & Best fit & $68\%$ limits\\
\hline
$\Omega_bh^2$ & $0.009471$ & $0.02191_{-0.01691}^{+0.002329}$ & $0.0223$ & $0.02235_{-0.000179}^{+0.000178}$ & $0.0214$ & $0.02148_{-0.00014}^{+0.00023}$ \\
$\Omega_ch^2$ & $0.1077$ & $0.1108_{-0.02752}^{+0.02003}$ & $0.1209$ & $0.124_{-0.0039}^{+0.0025}$ & $0.1206$ & $0.120_{-0.0012}^{+0.0011}$ \\
$100\theta$ & $1.136$ & $1.048_{-0.06127}^{+0.06034}$ & $1.041$ & $1.041_{-0.000373}^{+0.000375}$ & $1.041$ & $1.040_{-0.00028}^{+0.00029}$ \\
$\tau$ & $0.4306$ & $0.3863_{-0.3763}^{+0.4136}$ & $0.084$ & $0.07043_{-0.0176}^{+0.018}$ & $0.0128$ & $0.02146_{-0.01246}^{+0.00268}$ \\
$\text{ln}(10^{10}A_s)$ & $3.644$ & $3.2435_{-0.5435}^{+0.1714}$ & $3.1$ & $3.073_{-0.0344}^{+0.0351}$ & $2.953$ & $2.966_{-0.01722}^{+0.01068}$ \\
$n_s$ & $0.9346$ & $1.017_{-0.02786}^{+0.08255}$ & $0.9634$ & $0.9609_{-0.00518}^{+0.00512}$ & $0.9565$ & $0.9567_{-0.004}^{+0.00404}$ \\
$w$ & $-1.024$ & $-1.459_{-0.08525}^{+0.4580}$ & $-1.674$ & $-1.691_{-0.359}^{+0.318}$ & $-2.012$ & $-1.828_{-0.184}^{+0.078}$ \\
$\lambda$ & $0.0006674$ & $<0.005621$ & $0.0007646$ & $<0.001781$ & $0.00022$ & $<0.00013$ \\
\hline
$H_0$ & $99.04$ & $89.28_{-3.448}^{+10.65}$ & $87.25$ & $84.63_{-4.9}^{+15.4}$ & $98.95$ & $93.2_{-2.203}^{+6.67}$ \\
$\Omega_d$ & $0.8798$ & $0.8300_{-0.02780}^{+0.05207}$ & $0.8111$ & $0.7859_{-0.0275}^{+0.07}$ & $0.8543$ & $0.834_{-0.006}^{+0.024}$ \\
$\Omega_m$ & $0.1201$ & $0.1699_{-0.05207}^{+0.02780}$ & $0.1889$ & $0.2141_{-0.07}^{+0.0275}$ & $0.1457$ & $0.166_{-0.024}^{+0.006}$ \\
$\sigma_8$ & $1.203$ & $0.9302_{-0.1083}^{+0.1089}$ & $1.006$ & $0.9833_{-0.0636}^{+0.102}$ & $1.034$ & $0.9958_{-0.0217}^{+0.05}$ \\
Age/Gyr & $12.17$ & $13.67_{-2.018}^{+1.259}$ & $13.61$ & $13.71_{-0.176}^{+0.102}$ & $13.56$ & $13.59_{-0.0557}^{+0.0291}$ \\
\hline
\end{tabular}
\end{table*}

\end{document}